\documentclass{article}
\usepackage{spconf,amsmath,graphicx,hyperref}
\usepackage{tabularx}
\usepackage{arydshln}
\usepackage{booktabs}
\usepackage{enumitem}

\newcolumntype{Y}{>{\centering\arraybackslash}X}

\renewenvironment{thebibliography}[1]{
  \begin{oldthebibliography}{#1}
  \setlength{\itemsep}{0.em}
  \setlength{\parskip}{0.em}
}
{
  \end{oldthebibliography}
}

\title{SLM-SS: Speech Language Model for Generative Speech Separation}
\name{Tianhua Li$^1$, Chenda Li$^{1,2,\dagger}$, Wei Wang$^1$, Xin Zhou$^1$, Xihui Chen$^1$, Jianqing Gao$^3$, Yanmin Qian$^{1,2,\dagger}$\thanks{This work was supported in part by China STI 2030-Major Projects under Grant No. 2021ZD0201500, in part by China NSFC project under Grants No. U25A20409, and in part by SJTU Med-X (Medicine \& Engineering) Translational Research Grant (YG2025LC09).
 $^\dagger$ represents the corresponding author.}}

\address{$^1$Auditory Cognition and Computational Acoustics Lab
\\MoE Key Lab of Artificial Intelligence, AI Institute
\\School of Computer Science, Shanghai Jiao Tong University, Shanghai, China \\
$^2$VUI Labs\\
$^3$AI research Institute, iFLYTEK Company Limited, Hefei, Anhui, China}
\begin{document}
%
\maketitle
\begin{abstract}
Speech separation (SS) has advanced significantly with neural network-based methods, showing improved performance on signal-level metrics. However, these methods often struggle to maintain speech intelligibility in the separated signals, which can negatively affect the performance of downstream tasks such as speech recognition.
In this work, we propose SLM-SS, a novel approach that applies speech language models to SS, aiming to enhance the intelligibility and coherence of the separated signals. We frame SS as discrete multi-codebook sequence generation, using Encoder-Decoder models to map quantized speech mixtures to target tokens. In addition to the autoregressive modeling strategy, we introduce a non-autoregressive model to improve decoding efficiency for residual tokens. Experimental results on the \texttt{LibriMix} dataset demonstrate that our approach shows significantly better preservation of speech intelligibility, leading to improved linguistic consistency in a variety of downstream tasks compared to existing approaches. 
\footnote{Demo: \url{https://herobrinelth.github.io/slm-ss}}
\end{abstract}
\begin{keywords}
speech language model, speech separation, encodec, speech intelligibility
\end{keywords}
\vspace{-10pt}
\section{Introduction}
\vspace{-5pt}
\label{sec:intro}



Speech separation (SS) is a crucial task in speech processing, aiming to isolate individual speech sources from a mixture of overlapping signals, with applications in areas such as automatic speech recognition (ASR), speaker identification (SID), and hearing aids. Currently, SS has been addressed using discriminative methods~\cite{tasnet-luo2018, conv-tasnet, bsrnn-luo2023, sepformer-cem2021}, typically trained with objectives like scale-invariant signal-to-distortion ratio (SI-SDR). Although effective in terms of waveform reconstruction, these approaches often fail to preserve speech intelligibility, introducing distortions that degrade downstream tasks like ASR. In contrast, generative methods~\cite{sepdiff-chenbo,DiffSep-robin2023} explicitly model the data distribution and can produce more coherent outputs, but they are commonly limited by slow iterative decoding and the risk of hallucinating non-existent speech.

Recent advances in large language models (LLMs) have substantially improved a wide range of speech processing tasks by enabling tighter integration of acoustic and linguistic information through quantized or discrete speech tokens. VALL-E~\cite{valle} and Seed-TTS~\cite{seedtts-anastassiou2024} employ LMs over neural codec-based discrete tokens to generate high-quality speech in the text-to-speech (TTS) task. For target speaker extraction (TSE), TSELM~\cite{tselm} leverages discrete outputs from WavLM~\cite{wavlm} to incorporate target speaker information, while UniSep~\cite{unisep} models discrete sequences from SoundStream~\cite{soundstream} and uses prompt audio for conditional separation. In ASR, Seed-ASR~\cite{seedasr-bai2024} and FireRedASR~\cite{fireredasr-xu2025} integrate LLMs to achieve advanced performance. Furthermore, SepALM~\cite{sepalm} applies an LM for end-to-end denoising in speech separation, while SELM~\cite{SELM} and TokenSplit~\cite{tokensplit} further demonstrate the effectiveness of discrete token modeling for SE and multitask speech processing beyond generation and recognition, demonstrating the applicability of discrete token modeling beyond generation and recognition tasks.

\begin{figure*}[!ht]
    \centering
    \includegraphics[width=\textwidth]{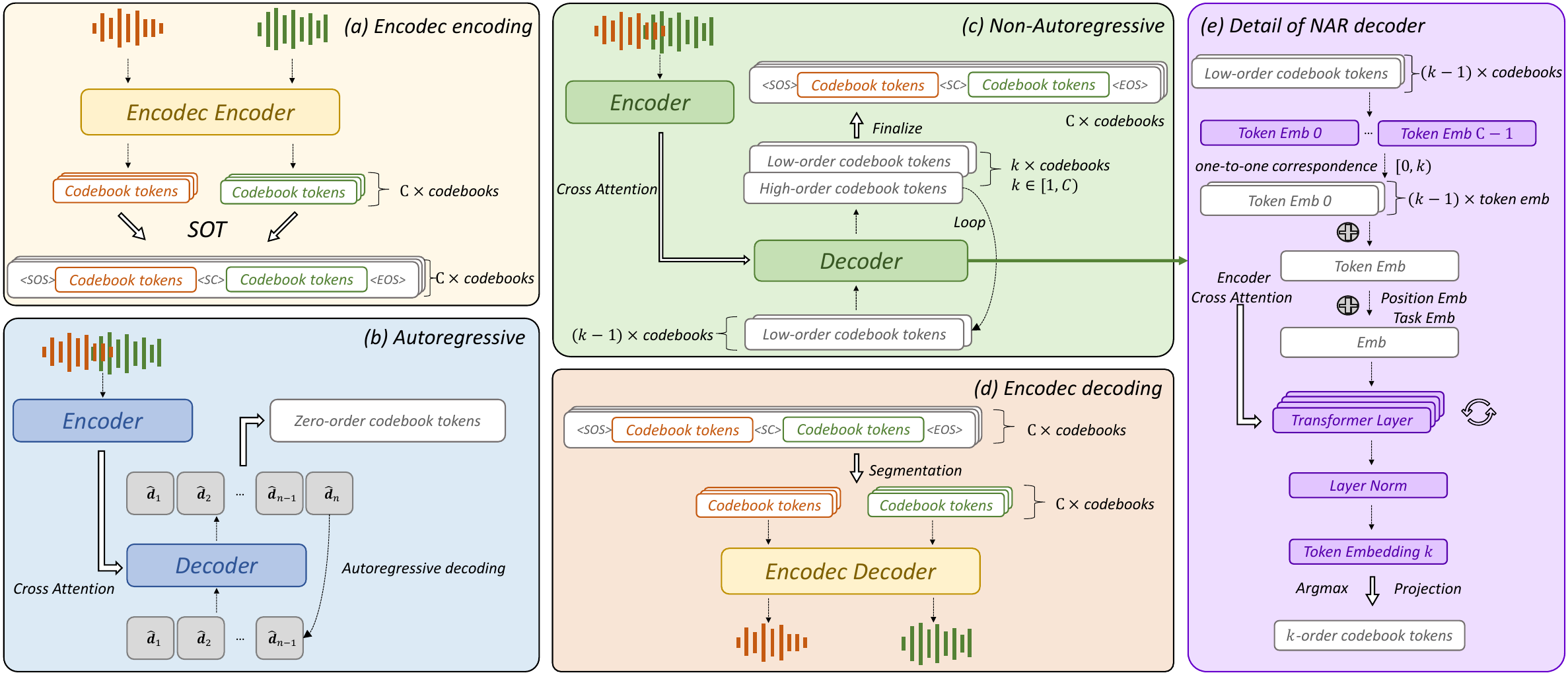}
    \caption{Overview of SLM-SS. (a) Encodec quantizes single-speaker audio into multi-codebook sequences then merges them using SOT. (b) AED model predicts the zero-order codebook sequence. (c) NAR model sequentially predicts higher-order codebook sequences given lower-order ones. (d) SOT sequences are segmented into single-person sequences then decoded into audio. (e) NAR decoder employs multiple independent token embeddings to integrate all low-order sequence information.}
    \label{fig:pipeline}

\end{figure*}

In this work, we introduce the SLM-SS framework, applying speech language models (SLMs) and quantized codecs to improve the intelligibility of separated speech. We employ Encodec~\cite{encodec} to convert speech into discrete codebook sequences. These sequences are concatenated using Serialized Output Training (SOT)~\cite{sot} inspired by current ASR work\cite{sotasr,sotasr2}, and modeled by a transformer-based autoregressive (AR) encoder-decoder, where the encoder extracts features from the mixture and the decoder aligns them with the discrete sequences through cross-attention. To further improve speech quality, we introduce a non-autoregressive (NAR) model that predicts higher-order codebooks from lower-order ones, improving decoding efficiency. Our framework offers a generative alternative for SS, achieving improved speech intelligibility and demonstrating outstanding performance on downstream tasks. 

Our contributions can be summarized as follows:
\begin{itemize}[topsep=0pt, itemsep=0pt]
\item We propose the SLM-SS framework, applying SLMs to the speech separation task.
\vspace{-5pt}
\item We utilize a hybrid AR and NAR generation scheme to improve speech quality while maintaining efficiency.
\vspace{-5pt}
\item We demonstrate through experiments that our approach significantly outperforms existing methods on downstream task performance.
\end{itemize}

\section{Method}
\label{sec:method}

\subsection{Generative Modeling of Speech Separation}
The flowchart of SLM-SS is shown in Fig. \ref{fig:pipeline}. Single-speaker audio is first encoded into multi-codebook sequences via Encodec, then concatenated using the SOT strategy to form multi-speaker sequences. The zero-order codebook is modeled with an AED framework, where the decoder performs autoregressive inference guided by cross-attention over multi-speaker features. Then, an NAR model with the same architecture predicts higher-order codebooks, using independent token embedding layers for each lower-order input and combining them to produce final embeddings. Finally, codebook sequence is segmented with special symbols and decoded by Encodec to yield separated single-speaker signals.

We employ Encodec to map continuous audio into compact discrete representations. Encoding produces 32-order codebooks with size $\mathcal{|C|}=1024$. Within the encoder-decoder framework, we extract and encode single-speaker segments from each multi-speaker clip provided by the dataset. Their multi-codebook sequences are then concatenated using the SOT strategy: a transcription start symbol \textit{$<$SOS$>$} is introduced, sequences are concatenated in a utterance first-in-first-out order, special separator symbols \textit{$<$SC$>$} denote speaker changes, and a termination symbol \textit{$<$EOS$>$} marks the end, namely:
\vspace{-5pt}
\begin{equation}
\begin{aligned}
\mathbf{C} &= [\mathbf{c}_0, \mathbf{c}_1, \dots, \mathbf{c}_{m-1}],\\
    \mathbf{c}_i&=[\text{SOS},r_{1,i}^1,\dots,r^1_{N^1,i},\text{SC},\,r_{1,i}^2,\dots,r^2_{N^2,i},\text{EOS}],
\end{aligned}
\end{equation}
where $r_{k,i}^j \in \mathcal{C}$ denotes the $k$-th token of the $j$-th speaker in $i$-th order codebook sequence, $N^j$ represents the sequence length corresponding to the $j$-th speaker, $m$ represents number of codebooks. It should be noted that even though all codebooks have the same vocabulary size and id space, the same numerical value may represent entirely different features in different codebooks, while special symbols have the same meaning in each codebook.

After obtaining the complete $m$ order SOT sequences from our SLM-SS method, we slice the multi-order sequences by \textit{$<$SC$>$} to derive single-person sequences. These sequences are then fed into the Encodec decoder to recover clean, single-person speech data. Since speaker transition is directly represented using special symbols, scenarios with an unknown number of speakers can be handled directly.
\vspace{-10pt}

\subsection{Hybrid Decoding for Progressive Codec Generation}

\subsubsection{AED Model for Zero-order Codebook Sequence}
For the encoder, we selected the pre-trained WavLM\cite{wavlm} model, a powerful Transformer-based feature encoder, and we chose to finetune on top of the pre-trained weights from WavLM. For the decoder, we follow the design of the Whisper decoder\cite{whisper} to build ours, and train it from scratch. The vocabulary $
\mathcal{V}$ encompasses the Encodec model's codebook size, along with three special symbols for SOT tasks, namely:
\begin{equation}
    \mathcal{V} = \mathcal{C} \cup \{\text{SOS},\text{SC},\text{EOS}\}.
\end{equation}

When obtaining the zero-order codebook sequence, the encoder first encodes multi-person audio into deep features. To integrate the capabilities of each hidden layer in WavLM, we designed a linear layer to fuse features from all layers. After layer normalization, the deep features $\mathbf{H}$ are fed into the decoder. Subsequently, the autoregression decoder predicts the $n$-th token $c_0^n$ based on the encoded audio features $\mathbf{H}$ by cross attention and the historical tokens $[c_0^1,\dots,c_0^{n-1}]$:

\begin{equation}
    \mathbf{o}_n = \text{Decoder}([c_0^1,\dots,c_0^{n-1}],\mathbf{H}),
\end{equation}
where $\mathbf{o}_n\in \mathbf{R}^{|\mathcal{V}|}$ is probability distribution of $n$-th token.

\vspace{-10pt}
\subsubsection{NAR Model for High-order Codebook Sequences}
We adopt the same model architecture as the AED model, but remove the unidirectional mask to implement an NAR model. As shown in Fig. \ref{fig:pipeline}(e), to achieve collaborative training of multi-order codebooks, we designed eight independent token embedding layers that share the same positional embedding. Additionally, we introduced task embedding to inform the model of the current codebook prediction task type. 

When predicting the $i$-th order codebook sequence, we must simultaneously consider information from all lower-order codebook sequences. This requires embedding and summing all lower-order sequences, then incorporating positional and task embeddings to obtain the total embedding $\mathbf{E}_i$, namely:
\vspace{-10pt}
\begin{equation}
\mathbf{E}_i = \Bigg( \sum_{j=0}^{i-1} \mathrm{Emb}\big(\mathbf{c}_j;\theta_j\big) \Bigg) 
      + \mathbf{P} + \mathbf{T}_i ,
\end{equation}
where $\mathrm{Emb}\big(\mathbf{c}_j; \theta_j \big), j\in[0,i)$ represents the token embedding of $j$-th codebook sequence, $\mathbf{P}$ is the positional embedding while $\mathbf{T}_i$ represents task embedding of $i$-th task. Then, $\mathbf{E}_i$ is subsequently fed into a series of Transformer layers for deep modeling and obtained $\mathbf{H}_i$. Finally, it is projected onto the $i$-th order codebook's token embedding $\mathbf{W}_i$ to yield the probability distribution $\mathbf{O}_i$ for all tokens in the $i$-th order codebook sequence:
\vspace{-5pt}
\begin{equation}
\begin{aligned}
\mathbf{H}_i &= \mathrm{Transformer}(\mathbf{E}_i), \\
\mathbf{O}_i &= \mathrm{Softmax}\!\left(\mathbf{H}_i \mathbf{W}_i^\top \right).
\end{aligned}
\end{equation}

\vspace{-15pt}
\section{Experiment}
\vspace{-5pt}
\label{sec:experiment}
\subsection{Experiment Setup}
\vspace{-5pt}
\textbf{Model}. The encoder is initialized with pre-trained WavLM-large weights, while the decoder only adopts dimensions from Whisper-medium, with adjusted vocabulary size and fewer (16) Transformer layers to meet resource limits, totally comprises about 600M parameters. Training runs for 30 epochs with an initial learning rate of $5\times 10^{-5}$, cosine annealing decay, and linear warm-up during the first 3 epochs. For AR beam search, we apply blank suppression, N-gram blocking  to avoid empty predictions and infinite repetition predictions. 

\noindent \textbf{Dataset \& Baseline.} Experiments are conducted on LibriMix \cite{librimix}, using 100h and 360h training subsets and evaluation on the test subset. Since Encodec’s encoding/decoding and partial codebook truncation may affect signal quality, we examine the relation between original audio and Encodec-processed versions. Although the original audio serves as the nominal groundtruth, our method is trained on 8-order Encodec codebooks rather than raw waveforms; thus, the effective upper bound is the audio reconstructed from 8-order codebooks. For comparison, we adopt BSRNN and Sepformer as baselines and apply Permutation Invariant Training (PIT) \cite{pit} to address the label permutation problem.

\noindent \textbf{Metric.} To evaluate perceptual quality, we conduct subjective listening tests with 20 volunteers on randomly sampled audio, where each model output is presented alongside the original recording in randomized order. Responses from a small number of participants who did not engage seriously were excluded. For linguistic consistency, we report word error rate (WER), Levenshtein Phoneme Similarity (LPS)\cite{lps}, and SpeechBERTScore (SBS) \cite{speechbertscore}. To further analyze Encodec’s internal consistency, we compute token error rate (TER), obtained from the zero-order codebook sequence after re-encoding the reconstructed audio.
\vspace{-15pt}
\subsection{Results Analysis}
\vspace{-5pt}
\begin{table*}[!htb]
    \centering
    \begin{tabularx}{\textwidth}{lYYYYYY}
        \toprule

        & \multicolumn{5}{c}{Speaker \& Linguistic Consistency} 
        & \multicolumn{1}{c}{Subjective} 
        \\
        \cmidrule(lr){2-6}  \cmidrule(lr){7-7}
        \textbf{Method}  & \textbf{Spk sim} &  \textbf{WER} & \textbf{TER} & \textbf{LPS} & \textbf{SBS} & \textbf{MOS}\\
        \midrule
       GT & - & 5.19 & - & 1.000 & 1.000 & 4.60\\
       GT-Encodec32 &  93.5 & 6.03 & 24.7 & 0.975 & 0.957  & 4.34\\
       \textbf{GT-Encodec8 (Upper Bound)} &  92.8 & 6.31 & 39.0 & 0.970 & 0.944 & 4.11\\
\cdashline{1-7}\noalign{\vskip\belowrulesep}
       BSRNN & \textbf{92.6} & 29.8 & 67.2 & 0.885 & 0.885 & 4.01\\
       Sepformer & 89.7 & 28.7 & 73.9 & 0.890 & 0.882 & 3.98\\
        \midrule
       \textbf{SLM-SS} & 91.7 & \textbf{7.24} & \textbf{45.8} & \textbf{0.954} & \textbf{0.913} & \textbf{4.19}\\
        \bottomrule
    \end{tabularx}
    \caption{Overall comparison of SLM-SS with existing approaches.}
    \label{table:result}
\end{table*}

Our experimental results are summarized in Table \ref{table:result}. Even when restoring the full 32-order codebook of Encodec, the generated audio still exhibits noticeable information loss, which becomes more pronounced when only the first 8 orders are used. This irreversible distortion, introduced by feature discretization, leads to slight degradation across objective metrics compared to the original audio, although it remains less perceptible in subjective listening tests and thus corroborates our hypothesis. A similar effect is reflected in TER: audio reconstructed from the 32-order codebooks still shows token errors after re-encoding, confirming the persistence of distortion. When decoding only the predicted 8-order codebooks before re-encoding, SLM-SS reveals stronger mismatches; however, the degradation remains less severe than that of Sepformer and BSRNN, owing to the Encodec-centered framework.

In contrast, SLM-SS introduces relatively mild distortion in terms of speaker identity and speech consistency. Our method demonstrates strong reconstruction ability, achieving word error rates (WER) close to the groundtruth. By comparison, BSRNN and Sepformer exhibit larger performance gaps in WER, which can be attributed to mismatches between pretrained ASR models and separation outputs—artifacts that are generally imperceptible to human listeners. A similar trend can be observed for LPS and SBS. 

Overall, both SLm-SS and the baseline models introduce mismatches and distortions, leading to inconsistent results across different evaluation metrics. To better capture their true impact, greater emphasis should be placed on subjective listening tests. In this regard, our method achieves consistently higher scores than BSRNN and Sepformer, while Encodec-reconstructed audio remains comparable to the original recordings. These findings confirm that our approach yields superior speech and feature reconstruction from a perceptual standpoint, where minor, imperceptible distortions can be safely disregarded.
\vspace{-10pt}
\subsection{Ablation on number of codebooks}
As mentioned earlier, our method is constrained by the number of codebooks. Table \ref{table:result} shows that restoration quality is strongly tied to codebook quantity—fewer codebooks yield more severe distortion. Theoretically, optimal results should be obtained by fully leveraging Encodec’s 32-order codebook sequence. However, since NAR models employ separate token embedding layers for different orders, predicting higher-order codebooks requires integrating information from all lower ones. This sharply increases training difficulty and computational cost, significantly slowing convergence. After comprehensive consideration of training costs and reconstruction quality, we ultimately selected the top 8 codebooks for modeling.

\begin{figure}[!htb]
    \centering
    \includegraphics[width=1\linewidth]{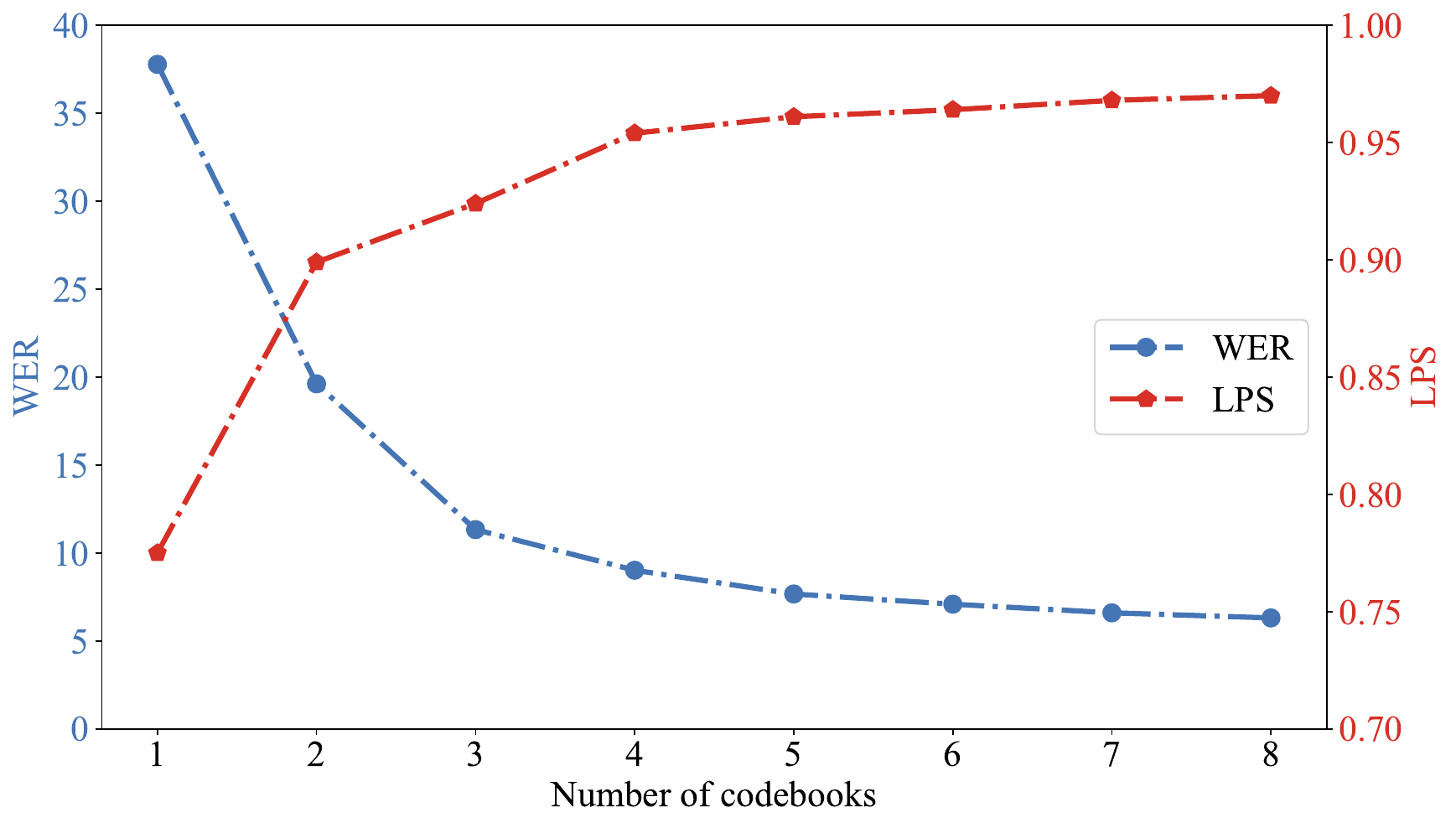}
    \vspace{-10pt}
    \caption{Variation of WER and LPS with Codebooks.}
    \label{fig:abla}
\end{figure}

To quantify this relationship, we evaluated several metrics with the first 1 to 7 codebooks and plotted performance trends in Fig. \ref{fig:abla}, where the performance shows a clear positive correlation as the number of codebooks increases. This demonstrates that furthermore increasing the number of codebooks can further enhance model performance.

\vspace{-10pt}
\subsection{Ablation on temperature during infernece}
In this experiment, we assess whether our method requires temperature tuning, as is often needed in models like VALL-E for performance gains. We conduct an ablation study on the temperature parameter during inference. The results in Table~\ref{table:temp} show that our method achieves optimal performance with the default temperature of 1.0, with no further tuning required. This indicates that our approach is less sensitive to temperature variations, making it easier to apply in practice.

\begin{table}[!htp]
    \centering
    \begin{tabular}{lccccc}
        \toprule
        \textbf{Temp.} & \textbf{Spk sim} & \textbf{WER} & \textbf{TER} & \textbf{LPS} & \textbf{SBS}\\
        \midrule
        0.5 & 38.9 & 49.1 & 69.3 & 0.581 & 0.695 \\
        0.9  & 73.1 & 10.2& 56.9& 0.900 & 0.845 \\
        1.0  & 91.7 & 7.2 & 45.8 & 0.954 & 0.913 \\
        1.1  & 77.8 & 9.7 & 52.0 & 0.949 & 0.895 \\
        1.5 & 54.2 & 64.6 & 87.8 & 0.178 & 0.497 \\
        \toprule
    \end{tabular}
    \vspace{-5pt}
    \caption{Performance on different AR temperature.}
    \label{table:temp}
\end{table}

\vspace{-15pt}
\section{Conclusion}
\label{sec:conclusion}
\vspace{-5pt}
This paper proposes SLM-SS, an SLM-based SS algorithm. After discretizing continuous speech signals into multi-codebook sequences via Encodec, we concatenate them according to the SOT strategy. We first employ an AR model to obtain the zero-order codebook, then use an NAR model to sequentially predict higher-order codebooks. The final multi-codebook sequence is sliced and decoded by Encodec to produce clean single-speaker audio. Extensive experiments demonstrate the effectiveness of our approach. Furthermore, we believe the ultimate goal lies in integrating SS and ASR tasks, enabling a single model to handle both. This direction will be explored in subsequent work.

\bibliographystyle{IEEEtran}
\bibliography{refs}

\end{document}